\documentclass[twocolumn,prb,showpacs,preprintnumbers,amsmath,amssymb,superscriptaddress]{revtex4}
\usepackage[dvipdfmx]{graphicx}
\usepackage{setspace}

\usepackage{color}
\usepackage{ulem} 

\usepackage[dvipdfmx,
colorlinks=true,
linkcolor=blue,
filecolor=blue,
urlcolor=blue]{hyperref}

\begin{document}
\title{Damping of the Collective Amplitude Mode in Superconductors with Strong Electron-Phonon Coupling}

\author{Yuta Murakami}
\affiliation{Department of Physics, University of Fribourg, 1700 Fribourg, Switzerland}
\author{Philipp Werner}
\affiliation{Department of Physics, University of Fribourg, 1700 Fribourg, Switzerland}
\author{Naoto Tsuji}
\affiliation{RIKEN Center for Emergent Matter Science (CEMS), Wako 351-0198, Japan}
\author{Hideo Aoki}
\affiliation{Department of Physics, University of Tokyo, Hongo, Tokyo 113-0033, Japan}
\affiliation{Electronics and Photonics Research Institute, 
Advanced Industrial Science and Technology (AIST), 
Umezono, Tsukuba, Ibaraki 305-8568, Japan}
\date{\today}

\begin{abstract}
We study the effect of strong electron-phonon interactions on the damping of the Higgs amplitude mode in superconductors by means of non-equilibrium dynamical mean-field simulations of the Holstein model. 
In contrast to the BCS dynamics, we find that the damping of the Higgs mode strongly depends on the temperature, becoming faster 
as the system approaches 
 the transition temperature.  
The damping at low temperatures is well described by a power-law, while near the transition temperature the damping shows exponential-like behavior.
We explain this crossover in terms of  
a temperature-dependent quasiparticle lifetime caused by the strong electron-phonon coupling, which smears the superconducting gap edge and makes the relaxation of the Higgs mode into quasiparticles 
more efficient at elevated temperatures. We also reveal that the phonon dynamics can soften the Higgs mode, which results in a slower damping.
\end{abstract}

\maketitle

\section{Introduction}

The problem of how a superconducting (SC) state evolves in time 
after an external 
stimulation 
has attracted the interest of researchers for a long time. 
\cite{Volkov1974,Littlewood1982,Barankov2004,Yuzbashyan2006a,Barankov2006,Yuzbashyan2006,Papenkort2007,Papenkort2008,Gurarie2009,Schnyder2011,Matsunaga2013,Zachmann2013,Matsunaga2014,Krull2014,Kemper2015,Sentef2015,Tsuji2015,Capone2015,Murakami2016,Krull2016} 
On the theoretical side, static mean-field analyses, which neglect inelastic collisions, have been widely employed to study the coherent dynamics. 
In the conventional weak-coupling BCS regime, coherent oscillations of the SC order parameter are known to 
decay with a power law ($\sim 1/\sqrt{t}$) regardless of temperature \cite{Volkov1974} and even beyond the linear-response regime. \cite{Yuzbashyan2006a,Barankov2006,Yuzbashyan2006,Papenkort2007,Papenkort2008,Zachmann2013} 
The power-law damping is remarkable, since it indicates a relatively slow decay without a specific relaxation timescale, 
which is usually observed only in special situations, such as near a critical point. 
Mean-field
analyses have further been applied to various situations to reveal changes in the properties of the coherent oscillations beyond the BCS regime for a bulk system.  
For instance, in the BEC regime the power law becomes $\sim 1/t^{1.5}$,\cite{Gurarie2009} while in quasi-1D systems the power strongly depends on the thickness of the system. \cite{Zachmann2013} The effect of a finite quasiparticle lifetime in the weak-coupling regime has been briefly 
addressed in Ref.~\onlinecite{Littlewood1982},
but the damping of the coherent oscillations in the correlated regime remains an interesting theoretical issue. 

The field has recently been stimulated by the observation of the collective amplitude (``Higgs")  mode \cite{Anderson1958,Higgs1964,Anderson1963} in conventional superconductors in pump-probe experiments with a strong THz laser pump.\cite{Matsunaga2013,Matsunaga2014} 
While previous experiments had already observed the collective amplitude mode in a coexistence region of superconductivity and charge order with Raman spectroscopy,\cite{Sooryakumar1980,Littlewood1981,Littlewood1982,Measson2014,Pekker2015}  the results reported in Refs.~\onlinecite{Matsunaga2013,Matsunaga2014} demonstrate the novel possibility of studying the collective mode in ordinary superconductors without coexisting orders. The pump-probe experiments can reveal not only the characteristic frequency of the collective mode but also its damping behavior after the pump. 
One important finding is that the damping of the coherent oscillations induced by a strong THz laser becomes significantly faster with increasing excitation 
intensity,\cite{Matsunaga2013} which is difficult to explain with the BCS dynamics without collisions.\cite{Yuzbashyan2006a,Barankov2006,Yuzbashyan2006,Papenkort2007,Papenkort2008,Zachmann2013} 
In addition, the temperature dependence of the damping is attracting experimental interest.\cite{Matsunaga_prep}
We also note that the sample used in these experiments, NbN, has a strong electron-phonon (el-ph) coupling.\cite{Wolf1985} 
It is therefore important 
to go beyond the BCS dynamics and to clarify the effects of strong el-ph couplings on properties of the amplitude Higgs mode.

Theoretical studies of the coherent oscillations in superconductors with strong el-ph couplings have only 
appeared 
recently, \cite{Kemper2015,Sentef2015,Murakami2016} and many important questions remain to be clarified.  
In this paper, we consider 
the effects of strong el-ph couplings on the temperature dependence of the damping behavior of the amplitude Higgs mode, taking into account the inelastic collisions of quasiparticles. 
Our study makes use of the non-equilibrium dynamical mean-field theory (DMFT), which enables us to simulate the damping behavior during the 
first several cycles 
after a pump pulse, as in pump-probe experiments. 
We find that the finite electron quasiparticle lifetime resulting from strong el-ph couplings leads to a strong temperature dependence of the damping of the amplitude mode.  
We also reveal that the phonon dynamics can soften the amplitude Higgs mode and extend its lifetime.

\section{Model and Method}
We focus on a basic model for el-ph coupled systems, the Holstein model, which is described by 
the Hamiltonian,
 \begin{equation}
 \begin{split}
 H(t)&=-\sum_{i,j,\sigma}v_{i,j}c_{i,\sigma}^{\dagger}c_{j,\sigma}-\mu\sum_i n_i+\omega_0\sum_i a^{\dagger}_i a_i\\
 &+g\sum_i (a_i^{\dagger}+a_i)(n_{i,\uparrow}+n_{i,\downarrow}-1),\label{eq:Holstein}
\end{split}
\end{equation}
where $c_{i,\sigma}^\dagger$ creates an electron with spin $\sigma$ at site $i$, $v_{i,j}$ is the electron hopping, $n_i=n_{i,\uparrow}+n_{i,\downarrow}$ 
with $n_{i,\sigma}=c_{i,\sigma}^\dagger c_{i,\sigma}$, and $\mu$ the 
electron chemical potential. $a_i^\dagger$ creates an Einstein phonon with a bare frequency $\omega_0$, and $g$ is the el-ph coupling.
We assume a semi-elliptic density of states for free electrons, $\rho(\epsilon)=\frac{1}{2\pi v^2_{\ast}} \sqrt{4v^2_\ast-\epsilon^2}$. We take $v_\ast=1$ as the unit of energy, and focus on the half-filled case ($\mu=0$). 
In this model, the phonon-mediated electron-electron attraction leads to an s-wave SC state, whose order parameter is $\phi=\frac{1}{N}\sum_i \langle c_{i\downarrow} c_{i\uparrow}\rangle$ with $N$ being the total number of sites.
We choose the order parameter to be real without loss of generality. 
In order to study the damping of the Higgs oscillations, i.e., the coherent oscillation of the amplitude of the SC order parameter, we consider a field coupled to the pair potential as a pump,  $F_{\rm ex}(t) \sum_i(c_{i\uparrow}^\dagger c_{i\downarrow}^\dagger+c_{i\downarrow}c_{i\uparrow})$. 
To be precise, we directly simulate the dynamics after a pump $F_{\rm ex}(t)=d_{\rm f}\delta(t)$ using the non-equilibrium DMFT framework (see below), with a small enough $d_{\rm f}$ (the linear response regime),
 and evaluate the dynamical pair susceptibility, \cite{Murakami2016}
\begin{equation}
 \chi_{\rm pair}(t-t')=-i\theta(t-t') \langle [B_{\bf 0}(t),B_{\bf 0}(t')] \rangle, \label{eq:pair_sus}
 \end{equation} 
 where $B_{\bf 0}=\sum_i(c_{i\uparrow}^\dagger c_{i\downarrow}^\dagger+c_{i\downarrow}c_{i\uparrow})$ and $\theta(t)$ is the step function.
Let us comment on a few points. (i) Even though a pump in the form of a pair potential field and the measurement of the dynamical pair susceptibility are rather academic tools, they allow us to focus on the amplitude dynamics of the order parameter, which has also been considered in previous investigations of the Higgs amplitude mode.\cite{Kulik1981,Benfatto2014,Benfatto2015}
 (ii) The pair potential field is related to more realistic excitations. For example, a modulation of the effective attractive interaction ($-\lambda$) can, within the BCS picture, 
 be regarded as a pair potential field pulse, 
 since the interaction term is $-\lambda (\langle c^\dagger_{\uparrow} c^\dagger_{\downarrow}\rangle c_{\downarrow}c_{\uparrow}+ c^\dagger_{\uparrow} c^\dagger_{\downarrow} \langle c_{\downarrow}c_{\uparrow}\rangle)$. We can expect the same effect for the modulation of the hopping parameter, since 
 by changing the measure of time, we can map the hopping modulation to an interaction modulation. 
 We also note that the hopping modulation can be realized as a second order effect of the electro-magnetic field \cite{Tsuji2011,Tsuji2015}, or by modulation of a certain phonon mode.\cite{Sentef2015,Subedi2011,Forst2011a}. 
 (iii) Our goal here is to evaluate the linear response function, Eq. (\ref{eq:pair_sus}), from a simulation of the time-evolution after a pump.
 In principle, one can obtain the same quantity by solving the Bethe-Salpeter equation. 
 However, the latter procedure usually involves a numerical analytic continuation, which introduces ambiguities.
 By calculating the real-time information directly,  we can avoid the analytic continuation.  
 (Another way to avoid solving the Bethe-Salpeter equation has recently been proposed in Ref.~\onlinecite{Tsuji2016}.)
 Even though we can only access the first several oscillations after the pump with the present approach, this is sufficient, since the pump-probe experiment also measure only the first several cycles after a pump.
 
The dynamics of the system is simulated using the framework of the non-equilibrium DMFT, \cite{Aoki2013} which becomes exact in the limit of infinite spatial dimensions.
In DMFT, the lattice model Eq.~(\ref{eq:Holstein}) is mapped onto a single-site impurity model, whose local Hamiltonian is $\mu n+\omega_0 a^\dagger a+g(a^\dagger+a)(n-1)$. 
The effective bath of the impurity problems is determined in a self-consistent manner such that the local electron Green's function $\hat{G}_{ii}(t,t')=-i\langle {\mathcal T}_{\mathcal C} \Psi_i(t) \Psi_i^\dagger(t') \rangle$ and the 
local 
self-energy $\hat{\Sigma}_{ii}$ of the lattice problem coincide with the impurity Green's function $\hat{G}_\text{imp}(t,t')=-i\langle {\mathcal T}_{\mathcal C} \Psi(t) \Psi^\dagger(t') \rangle$ and the impurity self-energy $\hat \Sigma_\text{imp}$, respectively. Here $\Psi^\dagger(t)\equiv[c_{\uparrow}^\dagger(t),c_{\downarrow}(t)]$ is a Nambu spinor, ${\mathcal T}_{\mathcal C}$ is the time ordering operator on the Kadanoff-Baym contour,  
and $\hat{\sigma}_{\alpha}$ a Pauli matrix, where a quantity with a 
hat represents a $2\times2$ Nambu-Gor'kov 
matrix.  
Similarly, the impurity phonon Green's function, $D_\text{imp}(t,t')=-i\langle {\mathcal T}_{\mathcal C} X(t)X(t')\rangle$, is identified with the local one in the lattice problem, $D_{ii}(t,t')=-i\langle {\mathcal T}_{\mathcal C} X_i(t)X_i(t')\rangle$. Here  $X=a^{\dagger}+a$.

We solve the non-equilibrium effective impurity problem with two types of diagrammatic approximations. 
The first is the self-consistent Migdal (sMig) approximation. \cite{Murakami2015,Murakami2016,Bauer2011,Freericks1994,Capone2003,Hague2008,Leeuwen2015,Pavlyukh2016} Here, the electron self-energy ($\hat{\Sigma}$) 
and phonon self-energy ($\Pi$) are expressed as 
\begin{equation}
\begin{split}
\hat{\Sigma}^{\rm sMig}(t,t')&=ig^2D_{\rm imp}(t,t')\hat{\sigma}_3\hat{G}_{\rm imp}(t,t')\hat{\sigma}_3,\\
\Pi^{\rm sMig}(t,t')&=-ig^2{\rm tr}[\hat{\sigma}_3 \hat{G}_{\rm imp}(t,t')\hat{\sigma}_3\hat{G}_{\rm imp}(t',t)].
\label{eq:Migdal}
\end{split}
\end{equation}
The sMig approximation neglects vertex corrections for the self-energies, which is justified when the phonon frequency is sufficiently smaller than the electron bandwidth. 
The dimensionless el-ph coupling is defined as $\lambda_{\rm eff}=-\rho(0)g^2 D^R(\omega=0)$. 
Since we are interested in the strong-coupling regime, we choose $\lambda_{\rm eff}\sim 1$.
With both self-energies considered self-consistently, electrons and phonons are renormalized, and their dynamics, including collisions between them, are taken into account.
The detailed diagrammatic expression for $\chi_{\rm pair}$ in this approximation has been presented in Ref.~\onlinecite{Murakami2016}. 
In addition to the ladder diagrams with electron legs, which are already taken into account in the BCS dynamics and include the effect of the relaxation of the Higgs mode into quasiparticle excitations, 
the self-consistent Migdal approximation includes the ladder diagrams with phonon legs and hybridizations between these two types of diagrams.

The other approximation is the unrenormalized Migdal (uMig) approximation,\cite{Murakami2015,Murakami2016,Kemper2015,Sentef2015} which corresponds to an electron self-energy
\begin{equation}
\hat{\Sigma}^{\rm uMig}(t,t')=ig^2D_0(t,t')\hat{\sigma}_3\hat{G}_{\rm imp}(t,t')\hat{\sigma}_3,
\end{equation} 
where $D_0$ is the bare phonon propagator and  the dimensionless el-ph coupling is defined as $\lambda_{\rm eff}=-\rho(0)g^2 D_0^R(\omega=0)$,
which we choose $\sim 1$ in this paper.
In this approximation, while electrons are renormalized and their collisions with phonons are considered, 
the phonons are not renormalized, stay in equilibrium and act as a heat bath.  
The diagrammatic expression for $\chi_{\rm pair}$ in the unrenormalized Migdal approximation 
contains the same 
type of diagrams as the BCS theory.\cite{Murakami2016} 
Thus the relaxation into quasiparticle excitations is included  in this approximation, but the coupling to the phonon dynamics and possible relaxation channels to phonons are ignored.
 Neglect of  the phonon renormalisation makes the approximation 
less accurate than the self-consistent Migdal approximation for describing the isolated Holstein model. \cite{Murakami2015} 
However, the uMig approximation  phenomenologically describes a situation where the phonons, which are equilibrated by other degrees of freedom than the focused system, act as a heat bath for the electrons (in an open system, beyond the pure Holstein model description). 
Considering this, we introduce a finite phonon lifetime $\Gamma$ 
in the unrenormalized Migdal approximation.
The phonon part is expressed as 
$D^R_0(\omega,\Gamma)=\frac{2\omega_0}{(\omega+i\Gamma)^2-\omega_0^2}$, 
where $R$ stands for the retarded part. 
The other components of the Green's function (lesser, greater etc.) are connected to the retarded part by the equilibrium fluctuation-dissipation theorem. \cite{Aoki2013}

In order to analyze the decay of the amplitude mode after a pump at $t=0$, we employ two types of fitting functions,
\begin{equation}
F_1(t)=a\;\exp(b\; t)+c\; \sin(\omega_{\rm H}\; t+d)/(t-t_0)^{\gamma}|_{t_0=0},\label{eq:power}\\
\end{equation}
\begin{equation}
F_2(t)=a\;\exp(b\; t)+c\; \sin(\omega_{\rm H}\; t+d)\exp(-\gamma \;t) \label{eq:exp},
\end{equation}
where $a, b, c, d, \gamma$ and $\omega_{\rm H}$ are fitting parameters.  
We use a least-square fit in the time interval $t\in[t_{\rm min},t_{\rm max}]$, where $t_{\rm min}$ is chosen as the first time at which $\chi_{\rm pair}(t)=0$.
 \begin{figure}[t]
  \centering
      
   \includegraphics[width=85mm]{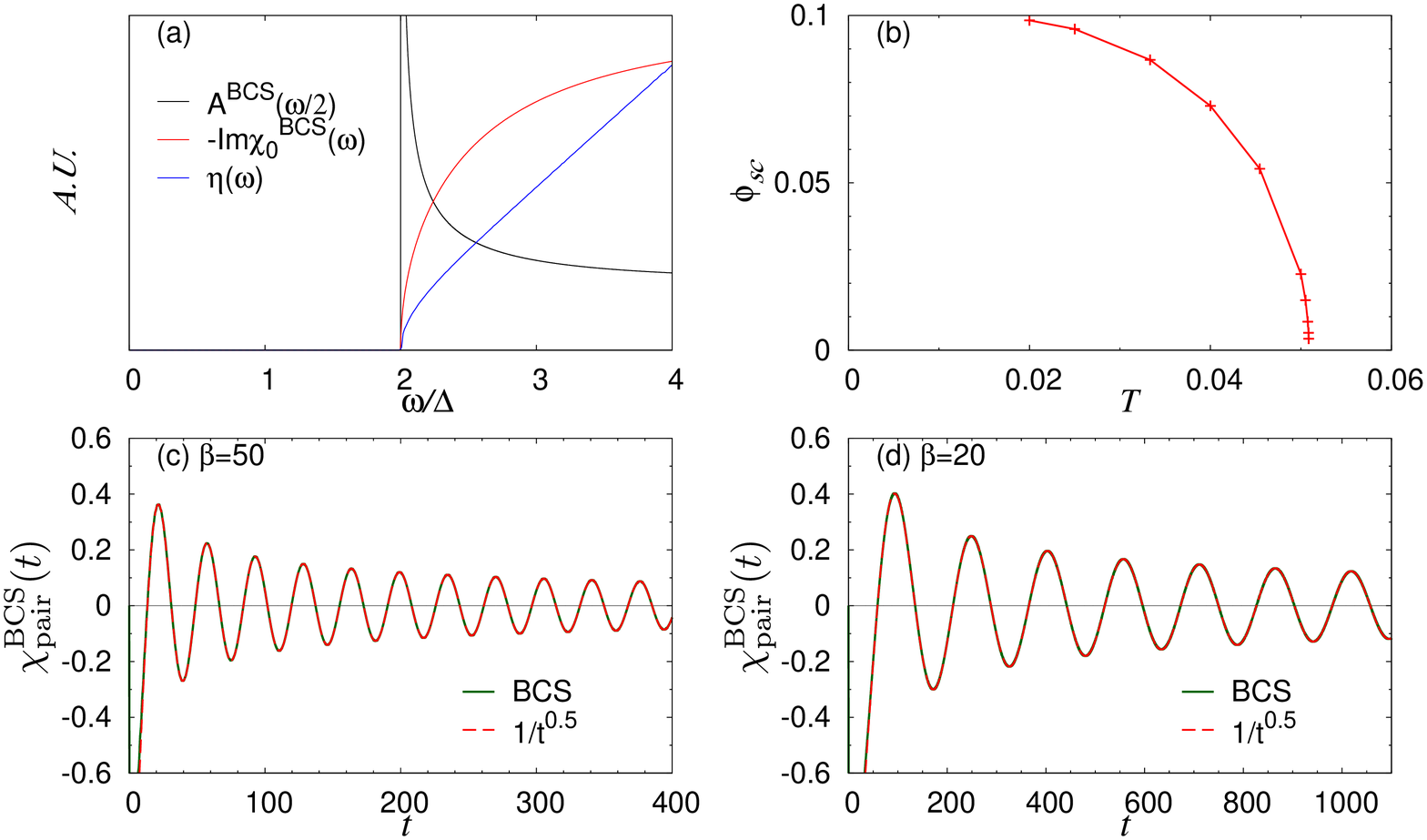}
  \caption{(a) BCS spectrum ($A^\text{BCS}(\omega/2)$), $-{\rm Im}\chi^{\rm BCS}_{\rm pair,0}(\omega)$ and $\eta(\omega)$ for $\lambda=0.9, \omega_c=2$ at $\beta=50$. (b) Temperature dependence of the order parameter. (c)(d)$\chi^{\rm BCS}_{\rm pair}(t)$ at $\beta=50$ (c) and $\beta=20$ (d). Red dashed lines, which almost completely overlap with the solid green lines, represent the result of the fitting function (\ref{eq:power}). 
  }
  \label{fig:chi_t_bcs2}
\end{figure}
\section{Results}

Since we want to clarify the difference to the BCS mean-field dynamics, we first recapitulate the dynamical pair susceptibility obtained within the BCS approximation.
We assume that the attractive interaction is represented as $\frac{1}{N}\sum_{{\bf k},{\bf k'}} V({\bf k},{\bf k'})c^\dagger_{{\bf k},\uparrow}c^\dagger_{-{\bf k},\downarrow}c_{-{\bf k'},\downarrow}c_{{\bf k'},\uparrow}$ with $V({\bf k},{\bf k'})=-\lambda\theta(\omega_c-|\epsilon_{\bf k}|)\theta(\omega_c-|\epsilon_{\bf k'}|)$, where $-\lambda$ represents the attractive interaction, $\omega_c$ is the cutoff energy and $\epsilon_{\bf k}$ is 
the bare electron dispersion with momentum ${\bf k}$.  
$\chi_{\rm pair}$ is expressed as 
\begin{equation}
\chi^{\rm BCS}_{\rm pair}(\omega)=\frac{\chi^{\rm BCS}_{\rm pair,0}(\omega)}{1+\lambda \chi^{\rm BCS}_{\rm pair,0}(\omega)/2}, \label{eq:BCS_chi_w}
\end{equation}
where the bubble contribution $\chi_{\rm pair,0}$ is the retarded part of $-i\frac{1}{N}\sum_{\bf k}\rm tr[\hat{\sigma}_1\hat{G}_{\bf k}(t,t')\hat{\sigma}_1\hat{G}_{\bf k}(t',t)]$. 
One can prove that ${\rm Re} \chi^{\rm BCS}_{\rm pair,0}(\omega)$ approaches $-2/\lambda$  linearly in the limit of $\omega\to 2\Delta_{\rm SC}+0$.
Here the SC gap in the BCS approximation is $\Delta_{\rm SC}\equiv-\lambda\frac{1}{N}\sum_{\bf k}\theta(\omega_c-|\epsilon_{\bf k}|)\langle c_{{-\bf k}\downarrow}c_{{\bf k}\uparrow}\rangle$.
The explicit expression of ${\rm Im} \chi^{\rm BCS}_{\rm pair,0}(\omega)$ is 
\begin{align}
&-\frac{1}{\pi}{\rm Im} \chi^{\rm BCS}_{\rm pair,0}(\omega)\nonumber\\
&=\int^{\omega_c}_{-\omega_c} d\epsilon \rho(\epsilon)\frac{\epsilon^2}{E^2}\tanh(\beta E/2)[\delta(\omega-2E)-\delta(\omega+2E)]\nonumber\\
&=\theta(|\omega|-2\Delta_{\rm SC})  \frac{2\rho(\kappa(\omega))\kappa(\omega)}{|\omega|} \tanh\Big(\frac{\omega\beta}{4}\Big),\label{eq:im_chi_bcs}
\end{align}
where $E=\sqrt{\epsilon^2+\Delta_{\rm SC}^2}$, and $\kappa(\omega)=\sqrt{\omega^2/4-\Delta_{\rm SC}^2}$. 
From this expression, one can see that $|{\rm Im} \chi^{\rm BCS}_{\rm pair,0}(\omega)|$ behaves  $(\omega-2\Delta_{\rm SC})^{1/2}$ in the vicinity of $\omega=2\Delta_{\rm SC}$, see Fig.~\ref{fig:chi_t_bcs2}(a). 
Hence the denominator of Eq.~(\ref{eq:BCS_chi_w}) becomes zero at $\omega=\omega_{\rm H}^{\rm BCS}=2\Delta_{\rm SC}$, which corresponds to the amplitude Higgs mode. 
We also note that if there were a pole at $\omega$, we could interpret the quantity $\eta(\omega)\equiv |{\rm Im} \chi_{\rm pair,0}(\omega)/(\frac{d {\rm Re}\chi_{\rm pair,0}(\omega)}{d\omega})|$ as the rate of the scattering from the collective mode to quasiparticle excitations, since it corresponds to the half-width of the peak in the spectrum.\cite{Littlewood1982} Therefore, Eq.~(\ref{eq:im_chi_bcs}) implies that below the SC gap ($|\omega|<2\Delta_{\rm SC}$) 
damping channels are energetically unavailable, while they are available above the gap ($|\omega|>2\Delta_{\rm SC}$). 
Because of the factor $(\epsilon/E)^2$ in the second line, the contribution from the energetically available channels becomes small near $\epsilon=0$. This results in a decrease of $|{\rm Im} \chi_{\rm pair,0}(\omega)|$ and $\eta(\omega)$ going to zero, see Fig.~\ref{fig:chi_t_bcs2}(a), which indicates a slower decay than an exponential decay of this mode. However, because of the rapid increase of $|{\rm Im} \chi_{\rm pair,0}(\omega)|$ and $\eta(\omega)$, this mode is not undamped.
Indeed, the analytic expression for the asymptotic behavior of  $\chi^{\rm BCS}_{\rm pair}$ is proportional to \cite{Volkov1974}
\begin{equation}
\begin{split}
\chi_{\rm pair}^{\rm BCS}(t)\propto\frac{1}{\sqrt{\Delta_{\rm SC}t}}\sin\Bigl(2\Delta_{\rm SC}t+\frac{\pi}{4}\Bigl).
\end{split}
\end{equation}
 In Fig.~\ref{fig:chi_t_bcs2}, we show $\chi^{\rm BCS}_{\rm pair}(t)$ along with the fitting function Eq.~(\ref{eq:power}). 
 Regardless of the temperature (both well below $T_c$ and near $T_c$), the fitting works very well {\it from the first oscillation} and the exponent is $1/2$. This has been found in previous works for various types of excitations.\cite{Volkov1974,Yuzbashyan2006,Papenkort2008}

 \begin{figure}[t]
  \centering
      \vspace{-0.3cm}
   \includegraphics[width=85mm]{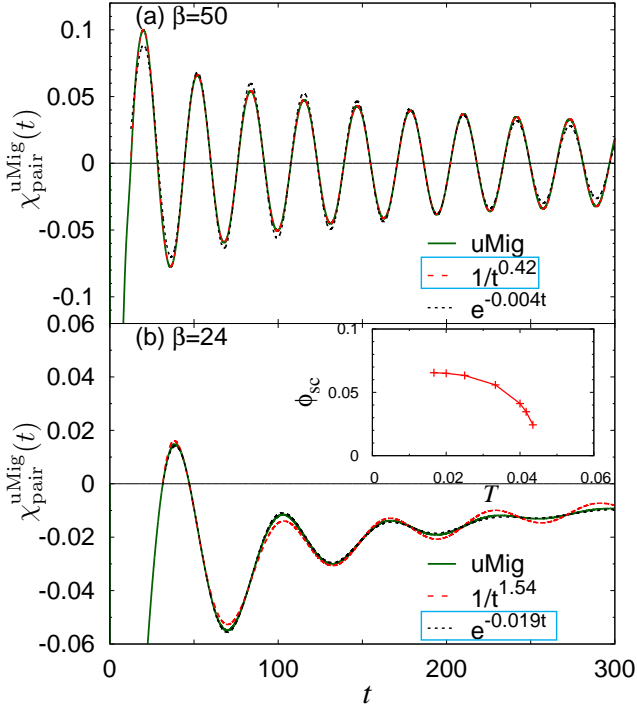}
  \caption{$\chi_{\rm pair}^{\rm uMig}(t)$ for $\omega_0=0.4,g=0.9,\Gamma=0.1$ ($\lambda_{\rm eff}\simeq1.2$) at $\beta = 50$ (a) and $\beta=24$ (b). 
 Here $t_{\rm max}=300$ is used for the fitting.
  The power-law and exponential fits are 
  shown as red dashed and blue dotted lines, respectively.  
Rectangles mark the better fits. The inset shows the temperature dependence of the order parameter.}
  \label{fig:fit_chi_pair_HF_w0.4g0.9gam0.1}
\end{figure}

Now we turn to the results of the unrenormalized Migdal approximation (with equilibrium phonons) in order to grasp the effects of the finite quasiparticle lifetime resulting from strong el-ph couplings.
Figure~\ref{fig:fit_chi_pair_HF_w0.4g0.9gam0.1} shows $\chi_{\rm pair}$ obtained within this scheme, which we denote by $\chi^{\rm uMig}_{\rm pair}(t)$. 
In contrast to the BCS dynamics, the damping exhibits a strong dependence on the temperature:  
When the temperature is much lower than $T_c$, the damping is well described by a power law but with a different exponent from the BCS value, while 
the exponential fitting is inadequate, see Fig.~\ref{fig:fit_chi_pair_HF_w0.4g0.9gam0.1}(a).
When the temperature is close to $T_c$, an exponential fit (Eq.~(\ref{eq:exp})) becomes better than a power-law fit (Eq.~(\ref{eq:power})), as in Fig.~\ref{fig:fit_chi_pair_HF_w0.4g0.9gam0.1}(b). 
In between these two regimes, neither of the two fitting forms 
can accurately describe the decay. 
We also note that the decay of the amplitude mode as discussed above is not limited to the present type of the pump protocol. 
The same damping behavior (not shown) is observed after a hopping modulation, $v(t)=v+\delta v\exp(-\frac{(t-t_{\rm pump})^2}{2\sigma_{\rm pump}^2})$, which mimics the effect of a strong laser. \cite{Murakami2016}
 \begin{figure}[t]
  \centering
   \includegraphics[width=85mm]{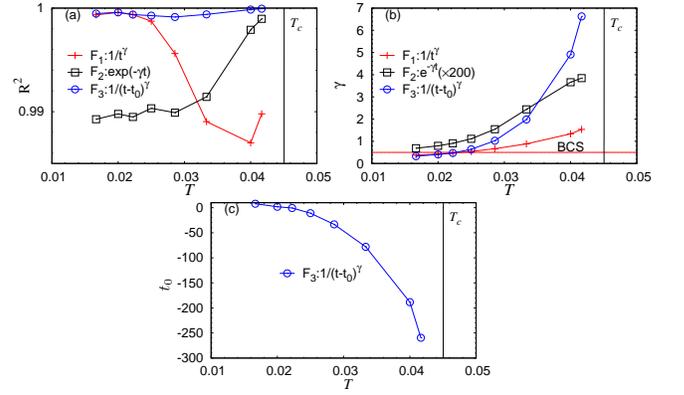}
  \caption{The coefficient of determination  $R^2$ (a) and the 
coefficient $\gamma$ (b) for each type of fitting for $g=0.9,\omega_0=0.4,\Gamma=0.1$. The vertical black lines indicate $T_c$, 
while the horizontal red line in panel (b) shows the BCS result (power law with $\gamma=1/2$).  (c) $t_0$ extracted from the $F_3$ fit.
}
  \label{fig:HF_detail}
\end{figure}


 \begin{figure}[t]
  \centering
   \vspace{0cm}
   \includegraphics[width=85mm]{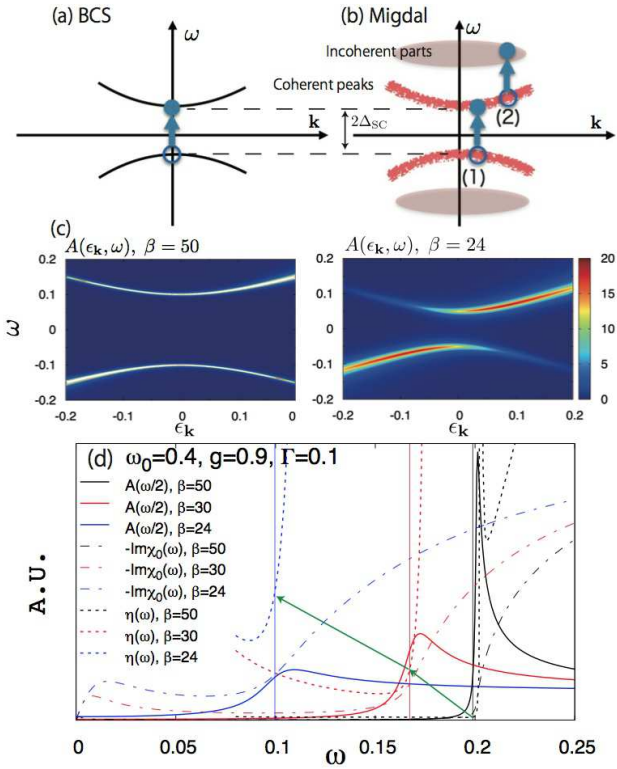}
  \caption{(a)(b) Schematic illustration of the difference between the BCS and Migdal descriptions of the excitations to which the amplitude mode decays. An arrow indicates an excitation process from an initial state of an electron (empty circle) to a final state (filled). Black curves in (a) represent the quasiparticle dispersion with an infinite lifetime, while red thick curves in (b) show the dispersion of the quasiparticles with a finite lifetime. Pink ovals in (b) represent the incoherent parts of the spectrum. 
  (c) $A(\epsilon_{\bf k},\omega)$ at $\beta=50$ (left) or $\beta=25$ (right) for $\omega_0=0.4,g=0.9,\Gamma=0.1$ within the unrenormalized Migdal approximation.
  (d) Comparison between the electron spectrum $A(\omega/2)$, $-{\rm Im}\chi^{\rm uMig}_{\rm pair,0}(\omega)$, $\eta(\omega)$ and $\omega_{\rm H}$ for $\omega_0=0.4,g=0.9, \Gamma=0.1$ at various temperatures. Vertical lines indicate the Higgs mode energy $\omega_{\rm H}$. For the green arrows, see text.}
  \label{fig:spectrum_Higgs_w0.4g0.9gam0.1}
\end{figure}

In Fig.~\ref{fig:HF_detail}(a), we show for each fitting function the 
coefficient of determination, 
which is defined for a set of data, $\{(t_i,y_i)\}$, and a fitting function $f(t)$ as $R^2\equiv 1-[\sum_i (y_i-f(t_i))^2]/[\sum_i (y_i-\bar{y})^2]$. Here $\bar{y}$ is the average of $y_i$, 
and a value of $R^2$ closer to $1$ indicates a better fit. 
It is evident that a power-law (Eq.~(\ref{eq:power})) provides the better description than an exponential law (Eq.~(\ref{eq:exp})) at low temperatures, while the opposite is true near $T_c$.
The damping coefficients are depicted in Fig.~\ref{fig:HF_detail}(b). In contrast to the BCS dynamics, the damping shows a significant dependence on temperature, i.e., the damping becomes faster with increasing temperature. 

Since in some previous analyses of the damping of the Higgs mode\cite{Zachmann2013,Matsunaga2013} $t_0$ in Eq. (\ref{eq:power}) was treated as a fitting parameter, we also consider this case and denote the corresponding fitting function as $F_3(t)$. We note that when $-t_0$ and $\gamma$ are large enough, $1/(t-t_0)^{\gamma}$ behaves $\propto\exp(\frac{\gamma}{t_0} t)$ for a finite range of $t$, so that 
$F_3$ can be regarded as an interpolation between $F_1$ and $F_2$.
The result, shown in Fig.~\ref{fig:HF_detail}, shows that 
$F_3$ always provides a better $R^2$ than $F_1$ and $F_2$. 
At low temperatures the fitted value of $t_0$ stays around zero which means that $F_3$ essentially behaves as $F_1$, while 
$-t_0$ and $\gamma$ increase in a similar manner with increasing temperature near $T_c$, which is consistent with an exponential behavior. 
We note that it is hard to provide a physical interpretation for large negative $t_0$ (long time before the pump) near $T_c$ and it would be more natural to consider that the fitting function $F_2$ is more essential there, even though $F_3(t)$ gives the best fit. We also note that in the BCS case $t_0$ always stays near zero regardless of the temperature.

 \begin{figure}[t]
  \vspace{-0.2cm}
  \centering
   \includegraphics[width=85mm]{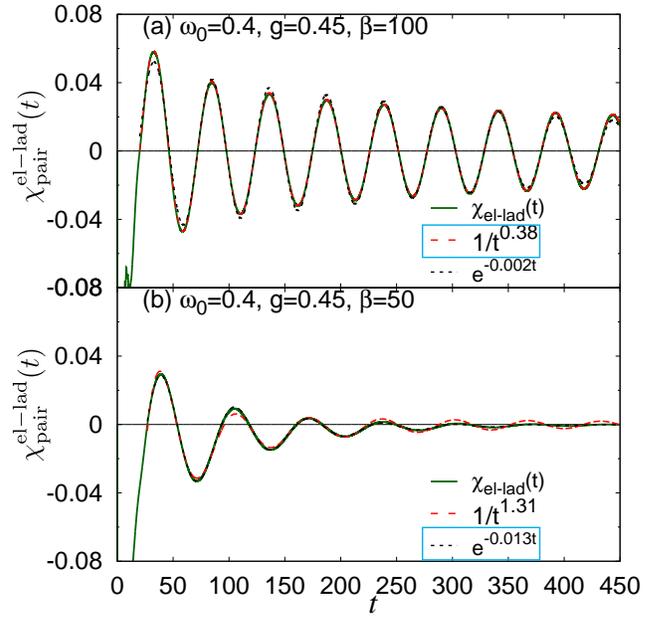}
  \caption{(a)(b) $\chi^{\rm el-lad}_{\rm pair}(t)$ for $\omega_0=0.4,g=0.45$  ($\lambda_{\rm eff}\simeq1.4$) at indicated temperatures. Here $t_{\rm max}=450$ is used for the fitting. 
  The power-law and exponential fits are also plotted. 
  Rectangles mark the better fitting function.} 
  \label{fig:Chi_pair_pair_el_lad_w0.4}
\end{figure}

Now we discuss the origin of the different decay behaviors in the BCS dynamics and the Migdal dynamics.
In the BCS theory, because of the relation $\omega_{\rm H}=2\Delta_{\rm SC}$ \cite{Volkov1974,Littlewood1982} and the fact that the lifetime of a quasiparticle is infinite, the relaxation channel of the amplitude mode (${\bf q}=0$) is limited to the quasiparticle excitations just at $\epsilon_{\bf k}=0$, see Fig.~\ref{fig:spectrum_Higgs_w0.4g0.9gam0.1}(a). 
However, as shown in Eq.~(\ref{eq:im_chi_bcs}), the contribution from this channel becomes 0 due to the factor $(\epsilon/E)^2$.
On the other hand, both Migdal approximations can take into account collisions, or the lifetime of the quasiparticles and the incoherent parts 
in the spectrum. 
In addition to this, as pointed out in our previous analysis,\cite{Murakami2016,Tsuji2016} the Higgs energy 
sticks to the SC gap ($2\Delta_{\rm SC}$) even for strong el-ph couplings. Here we emphasize that the relation between the Higgs energy and the SC gap is not trivial in the strong coupling regime and that it strongly affects the damping behavior of the collective mode.
Because of these situations, it becomes possible for the amplitude Higgs mode to decay into excitations from the lower band at various $\epsilon_{\bf k}\neq0$ ((1) in Fig.~\ref{fig:spectrum_Higgs_w0.4g0.9gam0.1}(b)), and into excitations to the incoherent parts ((2) in Fig.~\ref{fig:spectrum_Higgs_w0.4g0.9gam0.1}(b)). 
Since the energetically available relaxation channels are no 
longer restricted to $\epsilon_{\bf k}=0$, one can expect finite contributions from these channels. 
Now the quasiparticle lifetime decreases with increasing temperature, and the process (2) requires thermally excited quasiparticles above the Fermi energy. 
Hence these decay processes of the amplitude mode should become more significant closer to $T_c$ and make the decay faster.
We note that, in the present Holstein model, there is a phonon window, below which the quasiparticle lifetime becomes very long at low enough temperatures.
Therefore, at low enough temperatures the process (1) can {\it practically} use only $\epsilon_{\bf k}=0$, and this channel should suppressed as in the BCS case.
This situation should lead the distinct change of damping laws at low temperatures and temperatures around $T_c$.
In more realistic situations, one may need to consider acoustic phonons. In the presence of such phonons, the phonon window should become less clear and the quasiparticle lifetimes should decrease more quickly with increasing temperature. Hence it is expected that the damping laws at low temperatures and temperatures around $T_c$ becomes less distinct and that the damping tends to be faster and more exponential-like.
We also note that the temperature dependence of the damping 
has been briefly addressed in
Ref.~\onlinecite{Littlewood1982}, where they consider the effect of the quasiparticle lifetime on top of the BCS dynamics and 
suggest  
that the Higgs mode becomes overdamped near $T_c$.

The decrease of the quasiparticle lifetime is indeed evident in the temperature dependence of the electron spectrum, see Fig.~\ref{fig:spectrum_Higgs_w0.4g0.9gam0.1}(c)(d).
In Fig.~\ref{fig:spectrum_Higgs_w0.4g0.9gam0.1}(d), we show $A(\omega)=-\frac{1}{\pi} {\rm Im} G_{ii,11}(\omega)$ and the Higgs frequency derived from the fitting for various $T$, 
where $G_{ii,11}$ stands for the 11 component in the  Nambu-Gor'kov form.
The gap edge becomes smeared as we increase $T$, while the Higgs frequency is always located near the edge, i.e., $\omega_{\rm H}\simeq 2\Delta_{\rm SC}$ 
indeed holds.
Hence we conclude that near $T_c$ the relaxation of the amplitude mode into quasiparticles becomes efficient due to the strong el-ph coupling. This enhances the damping of the oscillations, which is well described by the exponential fit, Eq.~(\ref{eq:exp}). 
 Further support for this picture is obtained from $\eta(\omega)$, whose 
value at $\omega=\omega_H$ increases with increasing temperature, as shown by the green arrows in Fig.~\ref{fig:spectrum_Higgs_w0.4g0.9gam0.1}(d).\footnote{We note that ${\rm Im} \chi_{\rm pair,0}(\omega)$ changes smoothly. The rapid increase of $\eta(\omega)$ above $\omega_H$ comes from $\frac{d {\rm Re}\chi_{\rm pair,0}(\omega)}{d\omega}$. }
We note that, if the time dependence of the irreducible vertex can be approximately described by a delta function with a renormalized coefficient, a type of equation similar to Eq.~\ref{eq:BCS_chi_w} is 
 obtained, in which case we can indeed interpret $\eta(\omega)$ as the efficiency of the relaxation.

 \begin{figure}[t]
 \vspace{-0.2cm}
  \centering
   \includegraphics[width=85mm]{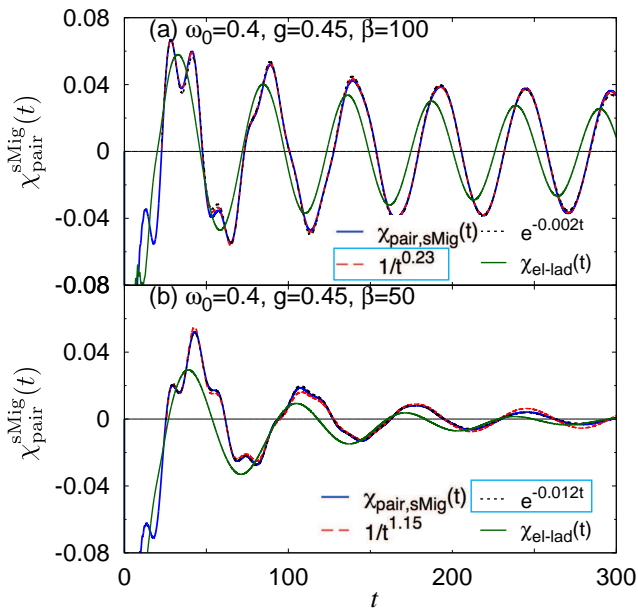}
  \caption{ (a)(b) Comparison between $\chi^{\rm sMig}_{\rm pair}(t)$ 
(along with fittings) and $\chi^{\rm el-lad}_{\rm pair}(t)$ for $\omega_0=0.4,g=0.45,\beta=100$ (a) and for $\beta=50$ (b). Here $t_{\rm max}=300$ is used for the fitting.
}
  \label{fig:Chi_pair_pair_direct_Mig_w0.4}
\end{figure}

Now, we move on to the self-consistent Migdal results, which include effects of the phonon dynamics. 
In order to single out the effects of the phonon dynamics on the Higgs mode, let us first look at the behavior of $\chi_{\rm pair}$ calculated with renormalized phonons but without phonon dynamics. Namely, we study the time evolution with the self-energy 
\begin{equation}
\hat{\Sigma}(t,t')=ig^2D^{\rm eq}_{\rm imp}(t,t')\hat{\sigma}_3\hat{G}_{\rm imp}(t,t')\hat{\sigma}_3,
\end{equation}
where the superscript ``eq" indicates that the propagator is the equilibrium one. 
We denote the pair susceptibility evaluated in this way as 
 $\chi^{\rm el-lad}_{\rm pair}$, since its diagrammatic expression consists of ladder diagrams with electron legs. \cite{Murakami2016}
 In the result shown in Fig.~\ref{fig:Chi_pair_pair_el_lad_w0.4}, 
the only difference from the uMig approximation is that the phonon propagator is renormalized through the el-ph coupling and depends on the temperature. \cite{Murakami2016}
As in the uMig case, at low temperatures a power-law fit, Eq.~(\ref{eq:power}), describes the damping well (see Fig.~\ref{fig:Chi_pair_pair_el_lad_w0.4}(a)), while around $T_c$ an exponential fit, Eq.~(\ref{eq:exp}), becomes 
better (see Fig.~\ref{fig:Chi_pair_pair_el_lad_w0.4}(b)). 
We note that, even when the phonon renormalisation is included, $\omega_{\rm H}$ sticks to the SC gap edge, $\omega_{\rm H}\simeq 2\Delta_{\rm SC}$, and 
the quasiparticle lifetime decreases with increasing temperature, which is the same as in the uMig approximation.  
In Fig.~\ref{fig:Chi_pair_pair_direct_Mig_w0.4}, we display 
$\chi^{\rm sMig}_{\rm pair}$ (along with fits) and 
$\chi^{\rm el-lad}_{\rm pair}$.
As pointed out in Ref.~\onlinecite{Murakami2016}, there emerges another collective amplitude mode originating from the phonon dynamics.
In order to deal with this, we have added a term $c' \sin(\omega' t+d')\exp(\gamma ' t)$ to the fitting functions (Eqs.~(\ref{eq:power}),(\ref{eq:exp})).
Again, the damping becomes faster with increasing temperature, and one can see a crossover of the damping from 
power law to exponential law. 
When we compare $\chi^{\rm sMig}_{\rm pair}$ and $\chi^{\rm el-lad}_{\rm pair}$, we notice different behavior between them, which indicates that $\chi^{\rm sMig}_{\rm pair}$ has a lower frequency \footnote{The difference is a few percent and $\omega_{\rm H}\simeq 2\Delta_{\rm SC}$ is valid regardless of the existence of the phonon dynamics.}, 
and that the oscillations in $\chi^{\rm sMig}_{\rm pair}$ are more slowly damped. Indeed, the fittings for $\chi^{\rm sMig}_{\rm pair}$ yield  smaller exponents than those for $\chi^{\rm el-lad}_{\rm pair}$, compare Fig.~\ref{fig:Chi_pair_pair_el_lad_w0.4} and Fig.~\ref{fig:Chi_pair_pair_direct_Mig_w0.4}. The softening of $\omega_{\rm H}$ can be attributed to the hybridization between the Higgs mode and the amplitude mode originating from phonon oscillations. 
The softening of the Higgs mode makes it longer-lived due to the suppression of the available relaxation channels to quasiparticles. 
Even though the possible decay of the Higgs mode into two phonons is also considered  within the self-consistent approximation, this channel is energetically suppressed by the reduction of the renormalized single-phonon spectral weight at $\omega \lesssim 2\Delta_{\rm SC}$(phonon anomaly). \cite{Allen1997,Murakami2016}

We finally comment on the relation between uMig and sMig. 
Firstly, recall that 
these two methods 
describe different physical set-ups: in sMig the system is isolated, while in uMig the system is open and the feedback from the phonon dynamics is neglected. 
Our results show that in both cases, one observes a crossover of the damping law as we vary the temperature. 
The common origin is the decreased quasiparticle lifetime and subsequent enhancement in the number of the available relaxation channels to quasiparticles with increasing the temperature. 
On the other hand, we would like to note that the uMig and sMig descriptions can lead to very different dynamics 
since they approach final states with different temperatures after strong excitations.\cite{Murakami2015}
In realistic situations, we may need to include both, the energy dissipation from the system and the feedback from the phonon dynamics. Which description is more appropriate depends on the specific problem.

\section{Conclusion}
We have studied the damping of the amplitude Higgs mode in a strongly-coupled phonon-mediated superconductor described by the Holstein model. The 
non-equilibrium DMFT results show that, in a sharp contrast to the BCS dynamics, the damping exhibits a strong temperature dependence and becomes faster as $T\rightarrow T_c$. 
Specifically, we have revealed that at low temperatures the damping of the Higgs oscillations is well described by a power law with an exponent distinct from the BCS value,
and that near the transition temperature the oscillations tend to decay with an exponential law.
In addition, we have shown that the phonon dynamics can soften the Higgs frequency and extend its lifetime. 

Our study has focused on the initial several cycles of the coherent oscillations after a pump. How the amplitude mode behaves in the long-time limit and how its damping is related to the behavior extracted here from the initial cycles are questions which should be clarified in the future. It is also important to understand the effects of el-ph couplings beyond the Holstein model, such as couplings with acoustic phonons, as well as the effects of the band structure, impurities and the system size.

\acknowledgments
 The authors wish to thank R. Matsunaga and R. Shimano for fruitful discussions and for showing us the experimental
result prior to publication.
HA is supported by a JSPS KAKENHI (No. 26247057) 
and ImPACT project (No. 2015-PM12-05-01) from JST, while YM has been supported by JSPS Research Fellowships for Young Scientists. PW acknowledges support from FP7 ERC starting grant No.~278023. NT is supported by JSPS KAKENHI (No.~25800192).

\bibliography{/Users/murakamiyuta/Documents/research/Reference/Ref}

\end{document}